\documentclass{iopart}  %% Normal journal article for NJP

\usepackage{cite}
\usepackage[]{graphicx}
\usepackage[percent]{overpic}
\usepackage[]{amssymb}
\usepackage{hyperref}
\begin{document}

\title{Coordinate transformation makes perfect invisibility cloak with arbitrary shape}

\author{Wei Yan, Min Yan, Zhichao Ruan, Min Qiu$^*$}%

\address{Laboratory of Optics, Photonics and Quantum Electronics,
  Department of Microelectronics and Applied Physics, Royal Institute
  of Technology, 164 40 Kista, Sweden}%
\ead{min@kth.se}
%\thanks{http://web.it.kth.se/~min/}

%\date{\today}
%\maketitle

\begin{abstract}
By investigating wave properties at cloak boundaries, invisibility
cloaks with arbitrary shape constructed by general coordinate
transformations are confirmed to be perfectly invisible to the
external incident wave. The differences between line transformed
cloaks and point transformed cloaks are discussed. The fields in the
cloak medium are found analytically to be related to the fields in
the original space via coordinate transformation functions. At the
exterior boundary of the cloak, it is shown that no reflection is
excited even though the permittivity and permeability do not always
have a perfect matched layer form. While at the inner boundary, no
reflection is excited either, and in particular no field can
penetrate into the cloaked region. However, for the inner boundary
of any line transformed cloak, the permittivity and permeability in
a specific tangential direction are always required to be infinitely
large. Furthermore, the field discontinuity at the inner boundary
always exists; the surface current is induced to make this
discontinuity self-consistent. For a point transformed cloak, it
does not experience such problems. The tangential fields at the
inner boundary are all zero, implying no field discontinuity exists.
\end{abstract}

 \submitto{\NJP} \maketitle

\section{Introduction}
The recent exciting development of invisibility cloaks, has
attracted intense attentions and discussions
\cite{Pendry}-\cite{Weber}. Theoretically, the cloaks are
constructed easily based on a coordinate transformation method as
proposed in Ref. [1]. The object inside the cloak is invisible to
the outside observer, because the light is excluded from the object
and the exterior field is not perturbed. The invisibility of the
linearly radially transformed cylindrical and spherical cloaks has
been confirmed by both numerical calculations \cite{Cummer,Zolla}
and analytical solutions \cite{Chen,Ruan}. Experimentally, the
invisibility cloak with simplified material parameters has been
implemented by Schurig et.al at the microwave frequency. Inspired by
the idea of the invisibility cloak, some interesting applications,
such as field concentration \cite{Rahm}, field rotation
\cite{Ychen}, and electromagnetic wormholes \cite{Greenleaf}, have
been proposed.

Up to now, most of discussions on invisibility cloaks focus on the
cylindrical and spherical cloaks produced by a coordinate
transformation only in the radial direction. For instance, in Ref.
[1], linearly radially transformed cylindrical and spherical cloaks
are discussed in detail, and their invisibility is confirmed by ray
tracing. In Refs. [5] and [6], the invisibility performances of such
cylindrical and spherical cloaks are further confirmed by obtaining
the exact fields in the cloak medium directly from Maxwell's
equations. In practice, it is sometimes desirable to have
invisibility cloaks whose shapes are tailored for the objects to be
cloaked. Thus, one needs to understand well about the properties of
invisibility cloaks with arbitrary shape produced by general
coordinate transformations. However, the investigations on an
invisibility cloak with arbitrary shape are only seen in few papers
\cite{U2,U21}. The mechanism why the invisibility of a general cloak
produced by compressing space is ensured, is still unclear. In this
paper, we investigate the electromagnetic (EM) properties of
invisibility cloaks with arbitrary shape constructed by general
coordinate transformations, and we confirm their perfect
invisibility. To figure out invisibility cloaks' main physical
properties, we only focus on the ideal case without considering the
practical implementation in this paper.

The paper is organized as follows. In section II, Maxwell's
equations in a curved coordinate system are derived. In section III,
we show how to construct an invisibility cloak by compressing space
in a general manner. In section IV, the wave behaviors and the
medium properties at the exterior boundary of the cloak are
investigated. In section V, we study the the wave behaviors and the
medium properties at the inner boundary of the cloak. Through
sections IV and V, the invisibility of cloaks with arbitrary shape
is confirmed, and the fields in the cloak medium are derived with
simple expressions. In section VI, the cloak parameters and the
fields in the cloak are derived when the transformed space is
described under arbitrary coordinate system. In section VII, two
examples of invisibility cloaks, i.e., cylindrical and spherical
invisibility cloaks are investigated. In section VII, the paper is
summarized.

\section{Maxwell's equations in a curved coordinate system}
Maxwell's Equations in a Cartesian ($x,y,z$) space take the form as
\begin{eqnarray}
\nabla  \times {\bf E} =  - \frac{{\partial {\bf B}}}{{\partial
t}},\quad \nabla \times {\bf H} = \frac{{\partial {\bf
D}}}{{\partial t}} + {\bf j},\quad \nabla \cdot {\bf D} = \rho
,\quad \nabla \cdot {\bf B} = 0,
\end{eqnarray}
with
\begin{eqnarray}
{\bf D} = \epsilon_0\overline{\overline \varepsilon }  \cdot {\bf
E},\quad {\bf B} = \mu_0\overline{\overline \mu }  \cdot {\bf H}.
\end{eqnarray}
Consider the transformation from cartesian space to an arbitrary
curved space described by coordinates ${ q_1,q_2,q_3}$ with
\begin{eqnarray}
x = f_1 (q_1 ,q_{2,} q_3 ),\quad y = f_2 (q_1 ,q_{2,} q_3 ),\quad z
= f_3 (q_1 ,q_{2,} q_3 ).
\end{eqnarray}
The length of a line element in the transformed space is given by $
dl^2 = [dq_1 ,dq_2 ,dq_3 ]Q[dq_1 ,dq_2 ,dq_3 ]^T $, where the
superscript $"T"$ denotes the transpose of matrix, and $Q=gg^T$ with
\begin{eqnarray}
g = \left[ {\begin{array}{*{20}c}
   {\frac{{\partial f_1 }}{{\partial q_1 }}} & {\frac{{\partial f_2 }}{{\partial q_1 }}} & {\frac{{\partial f_3 }}{{\partial q_1 }}}  \\
   {\frac{{\partial f_1 }}{{\partial q_2 }}} & {\frac{{\partial f_2 }}{{\partial q_2 }}} & {\frac{{\partial f_3 }}{{\partial q_2 }}}  \\
   {\frac{{\partial f_1 }}{{\partial q_3 }}} & {\frac{{\partial f_2 }}{{\partial q_3 }}} & {\frac{{\partial f_3 }}{{\partial q_3 }}}  \\
\end{array}} \right].
\end{eqnarray}
The volume of a space element is expressed as $dv = \det(g)dq_1 dq_2
dq_3$, where $\det (g)$ represents the determinant of $g$. Here, it
is noted that the way of describing the space transformation in this
paper is similar as in Ref. [13], where the time transformation is
also taken into account. The space-time metric tensor ${\bf
g_{\alpha\beta}}$ defined in Ref. [13] is ${ {\bf
g_{\alpha\beta}}}=diag[1,-Q]$ in the present paper, where only space
transformation is considered.

Then Maxwell's equations in the curved space take the form as
\cite{Pendry, U2}
\begin{eqnarray}
\nabla _q  \times \widehat {\bf E} =  - \frac{{\partial \widehat
{\bf B}}}{{\partial t}},\quad \nabla _q  \times \widehat {\bf H} =
\frac{{\partial \widehat {\bf D}}}{{\partial t}} + \widehat {\bf j}
,\quad \nabla_q \cdot \widehat {\bf D} = \widehat\rho ,\quad
\nabla_q  \cdot \widehat {\bf B} = 0
\end{eqnarray}
with
\begin{eqnarray}
\widehat {\bf D} = \epsilon_0\widehat{\overline{\overline
\varepsilon } } \cdot \widehat {\bf E},\quad \widehat {\bf B} =
\mu_0\widehat{\overline{\overline \mu} } \cdot \widehat {\bf H},\\
\widehat{\overline{\overline \varepsilon } } = { \det(g)(g^T)^{ - 1}
}\overline{\overline \varepsilon } g^{ - 1} ,\quad
\widehat{\overline{\overline \mu } } = {\det(g)}{{(g^T)^{ - 1}
}}\overline{\overline \mu } g^{ - 1},\\
\widehat {\bf j} = \det(g)(g^T)^{-1} {\bf j},\quad
\widehat\rho  = \det(g)\rho,\\
\widehat {\bf E} = g{\bf E},\quad \widehat {\bf H} = g{\bf H},
\end{eqnarray}
where the superscript "-1" denotes the inverse of matrix.

The permittivity and permeability ${\overline{\overline \varepsilon
} }$ and ${\overline{\overline \mu } }$ in the Cartesian space are
considered for a general case, i.e., they can be tensors. It is seen
above that Maxwell's equations in the curved space have the same
form as in the Cartesian space. However, the definitions of the
permittivity, permeability, current density, and electric charge
density are different, as shown in Eqs. (7) and (8).
\section{Construction of invisibility cloaks}
To construct a cloak, one usually starts from compressing an
enclosed space with the exterior boundary unchanged \cite{Pendry}.
As seen in Fig. 1, the region enclosed by boundary $S_1$ is
compressed to the region bounded by the exterior boundary $S_1$ and
the interior boundary $S_2$. Such a space compression can be viewed
as a certain coordinate transformation described by Eq. (3), which
makes a connection between the points in the compressed space with
coordinates $(q_1,q_2,q_3)$ and the points with Cartesian
coordinates $(x,y,z)$ in the original space. The exterior boundary
$S_1$ satisfies $q_1=x,q_2=y,q_3=z$. Notice the interior boundary
$S_2$ is obtained by blowing up a line or a point \cite{G2}. Thus
the cloaks can be divided into two classes: line transformed cloaks
and point transformed cloaks. The compressed shaded region in Fig. 1
is the desired cloak.
\begin{figure}[htbp]
\centering
\includegraphics[width=5cm]{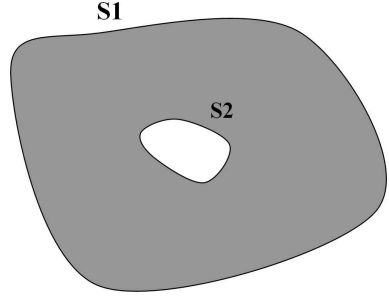}
\caption{The cross section of the cloak (shaded region). }
\end{figure}

The permittivity and permeability tensors of the cloak in Cartesian
coordinate system are given in Eq. (7). It seems that the cloak
medium is very complex, whose permittivity and permeability are
tensors and their values vary with the spatial location. However,
the eigen functions of the wave equations in the cloak medium are
quite simple, which relate with the eigen function of the
uncompressed space by Eq. (9).
%It is noted that the transformed compressed
%space is a virtual space, which should be transformed to the
%physical space, i.e., the Cartesian space. Therefore, $q_1,q_2,q_3$
%should be considered as $x,y,z$, respectively.

In order to achieve the invisibility, the cloak should be able to
exclude light from a protected object without perturbing the
exterior field. Thus for the above cloak, at the exterior boundary
$S_1$, external incident light should excite no reflection. While at
the interior boundary $S_2$, no reflection is either excited, and
light can't penetrate into the cloaked region. In the following
sections, we will prove the invisibility of the cloak by
investigating the wave behaviors at the cloak's exterior and inner
boundaries. For the simplicity of our discussions and considering
the practical application, the invisibility cloak is considered to
be placed in air. Then the permittivity and permeability of the
cloak in Eq. (7) will be simplified to
\begin{equation}
\widehat{\overline{\overline \varepsilon } } =
\widehat{\overline{\overline \mu } } = {\det (g)}(g^T)^{ - 1}g^{-1}
,
\end{equation}
which is the same as proposed in Refs. [1] and [13]. The cloak is
considered to be lossless at the working frequency.
\section{Cloak's exterior boundary}
In this section, we will prove that no reflection is excited at the
exterior boundary. The transmitted electric field $\widehat {\bf
E}^i$ and magnetic field $\widehat {\bf H}^i$ without interacting
with the inner boundary $S_2$ are expressed as
\begin{equation}
\label{1}
 \widehat {\bf E}^i  = g{\bf E}^i, \; \widehat {\bf H}^i  =
g{\bf H}^i,
\end{equation}
where ${\bf E}^i$ and ${\bf H}^i$ represent the electric and
magnetic fields of the external incident waves. According to Eq.
(9), it is easily seen that the fields expressed in Eq. (11) satisfy
Maxwell's equations in the cloak medium. Thus in order to prove no
reflection excited at $S_1$, one only needs to confirm that
tangential components of ${\bf E}^i$ (${\bf H}^i$) and $\widehat
{\bf E}^i$ ($\widehat {\bf H}^i$) keep continuous across $S_1$.

Decompose $\widehat{\bf E}^i$ and $\widehat{\bf H}^i$ into $
\widehat {\bf E}^i  = [\widehat E_n ^i ,\;\widehat E_{t_1}
^i,\;\widehat E_{t_2} ^i ]$ and $ \widehat {\bf H}^i  = [\widehat
H_n ^i ,\;\widehat H_{t_1} ^i,\;\widehat H_{t_2} ^i]$, where the
subscripts $"n"$ represent $S_1$'s normal direction pointing outward
from the cloak; $"t_1"$ and $"t_2"$ represent $S_1$'s two tangential
directions, which are vertical with each other. Thus Eq. (11) can
also expressed as
\begin{equation}
\label{3}
 \left[ {\begin{array}{*{20}c}
   {\widehat E_n ^i }  \\
   {\widehat E_{t_1} ^i }  \\
   {\widehat E_{t_2} ^i }  \\
\end{array}} \right] = [\widehat n,\;\widehat t_1,\;\widehat t_2 ]^{ - 1}
g{\bf E^i},\; \left[ {\begin{array}{*{20}c}
   {\widehat H_n ^i }  \\
   {\widehat H_{t_1} ^i }  \\
   {\widehat H_{t_2}^i} \\
\end{array}} \right] = [\widehat n,\;\widehat t_1,\;\widehat t_2 ]^{ - 1}
g{\bf H^i},
\end{equation}
where $\widehat n$, $\widehat t_1$, $\widehat t_2$ represent the
unit vectors in $n$, $t_1$, and $t_2$ directions, respectively.

At the exterior boundary $S_1$, $q_1=x$, $q_2=y$, and $q_3=z$. So
$f_i(q_1,q_2,q_3)-q_i=0$ ($i=1,2,3$) characterize the exterior
boundary $S_1$. Therefore, it is obvious that the vectors $\nabla_q
f_i- \widehat C_i$ ($i=1,2,3$) lie in the same line as the normal
direction $n$ of $S_1$, where $\widehat C_1=\widehat x$, $\widehat
C_2=\widehat y$, and $\widehat C_3=\widehat z$.  For the special
case when $\nabla_q f_i- \widehat C_i=0$, $\nabla_q f_i- \widehat
C_i$ can be expressed as $0\widehat n$, i.e., the vector with the
magnitude $0$ in the $n$ direction. Therefore, $g$ on $S_1$ can be
expressed as
\begin{equation}
g=[F_1\widehat n+\widehat x,\; F_2\widehat n+\widehat y,\;\widehat
F_3\widehat n+\widehat z],
\end{equation}
with
\begin{equation}
 |F_i|  = \sqrt {(\frac{{\partial f_i }}{{\partial q_i }} - 1)^2 +
(\frac{{\partial f_i }}{{\partial q_j }})^2  + (\frac{{\partial f_i
}}{{\partial q_k}})^2},
\end{equation}
where $ i,j,k = 1,2,3$ and $ i \ne j\ne k$; $F_i=|F_i|$ when the
direction of $\nabla_q f_i- \widehat C_i$ is as the same as the $n$
direction, and $F_i=-|F_i|$ if the direction of $\nabla_q f_i-
\widehat C_i$ is opposite to the $n$ direction. Substituting Eq.
(13) into Eq. (12) and noticing that $\widehat n$, $\widehat t_1$,
and $\widehat t_2$ are orthogonal with each other, it is easily
obtained that at $S_1$
\begin{equation}
\widehat { E}_{t_1} ^i  = {\bf E}^i  \cdot \widehat t_1,\;\widehat
{H}_{t_1} ^i  ={\bf H}^i  \cdot \widehat {t_1},
\end{equation}
\begin{equation}
\widehat { E}_{t_2} ^i  = {\bf E}^i  \cdot \widehat t_2,\;\widehat {
H}_{t_2} ^i  ={\bf H}^i  \cdot \widehat {t_2},
\end{equation}
which indicates that the tangential components of ${\bf E}^i$ (${\bf
H}^i$) and $\widehat {\bf E}^i$ ($\widehat {\bf H}^i$) are
continuous across $S_1$. Thus, it is proved that no reflection is
excited at the exterior boundary.

Consider the permittivity and permeability at $S_1$ for the
transformed cloak. It should be noticed that no flection excited at
the exterior boundary does not imply that the exterior boundary is a
perfectly matched layer (PML), where the permittivity and
permeability at $S_1$ have the PML form
$\widehat{\overline{\overline \varepsilon }
}=\widehat{\overline{\overline \mu }}=diag[u,1/u,1/u]$ with the
principle axes in $n$, $t_1$ and $t_2$ directions, respectively. We
find that parameters at $S_1$ have the PML form only when $g$ is a
symmetry matrix. Observing Eq. (13), we have $g^T\widehat
t_1=\widehat t_1$ and $g^T\widehat t_2=\widehat t_2$, indicating
that $\widehat t_1$ and $\widehat t_2$ are the eigen vectors of
$g^T$ with the same eigen value $1$. Considering $g$ is a symmetry
matrix, we can know the other eigen vector of $g^T$ is $\widehat n$
with eigen value $\det(g)$. Thus, $\widehat n$, $\widehat t_1$, and
$\widehat t_2$ are eigen vectors of $Q^{-1}=(g^T)^{-1}g^{-1}$, with
eigen values $1/\det(g)^2$, $1$, and $1$, respectively. Thus based
on Eq. (10), $\widehat{\overline{\overline \varepsilon } }$ and
$\widehat{\overline{\overline \mu } }$ for a symmetric $g$ can be
expressed as
\begin{equation}
\widehat{\overline{\overline \varepsilon } } =
\widehat{\overline{\overline u} } = diag[\frac{1}{{{\det (g)} }},\;
{\det (g)} ,\; {\det (g)} ],
\end{equation}
where the diagonal elements correspond to the principle axes
$\widehat n$, $\widehat t_1$, and $\widehat t_2$, respectively. The
radially transformed cylindrical and spherical cloaks fall into this
category\cite{Pendry}.

\section{Cloak's inner boundary}
In this section, we will prove that at the inner boundary $S_2$, no
reflection is excited and no field can penetrate into the cloaked
region. As discussed in section III, the inner boundary is
constructed by blowing up a line or a point, as seen in Fig. 2(a)
and (b). So in the following, two cases that: (1) line transformed
cloaks, (2) point transformed cloaks, will be discussed separately.

\subsection{ Case (1): line transformed cloaks}

Assume that $x=b_1(s)$ , $y=b_2(s)$ and $z=b_3(s)$ characterize the
line, which is mapped to the inner boundary $S_2$. We have
$f_1(q_1,q_2,q_3)=b_1(s)$ and $f_2(q_1,q_2,q_3)=b_2(s)$, and
$f_3(q_1,q_2,q_3)=b_3(s)$ at $S_2$. Each point $(b_1,b_2,b_3)$ on
the line maps to a closed curve on $S_2$. The parameter $s$ can be
expressed as a function of $q_1,q_2,q_3$ with $s=u(q_1,q_2,q_3)$.
$\nabla_q s=\partial u /\partial q_1 \widehat x+\partial u /\partial
q_2 \widehat y+\partial u /\partial q_3 \widehat z$ is the gradient
of $s$, which points in the direction of the greatest increase rate
of $s$. For $\nabla_q b_i$, we have $\nabla_q b_i=
\partial b_i /\partial s \nabla_q s$, where $i=1,2,3$. Thus,
$\nabla_q b_i$ and $\nabla_q s$ have the same direction.
\begin{figure}[htbp] \centering
\includegraphics[width=6cm]{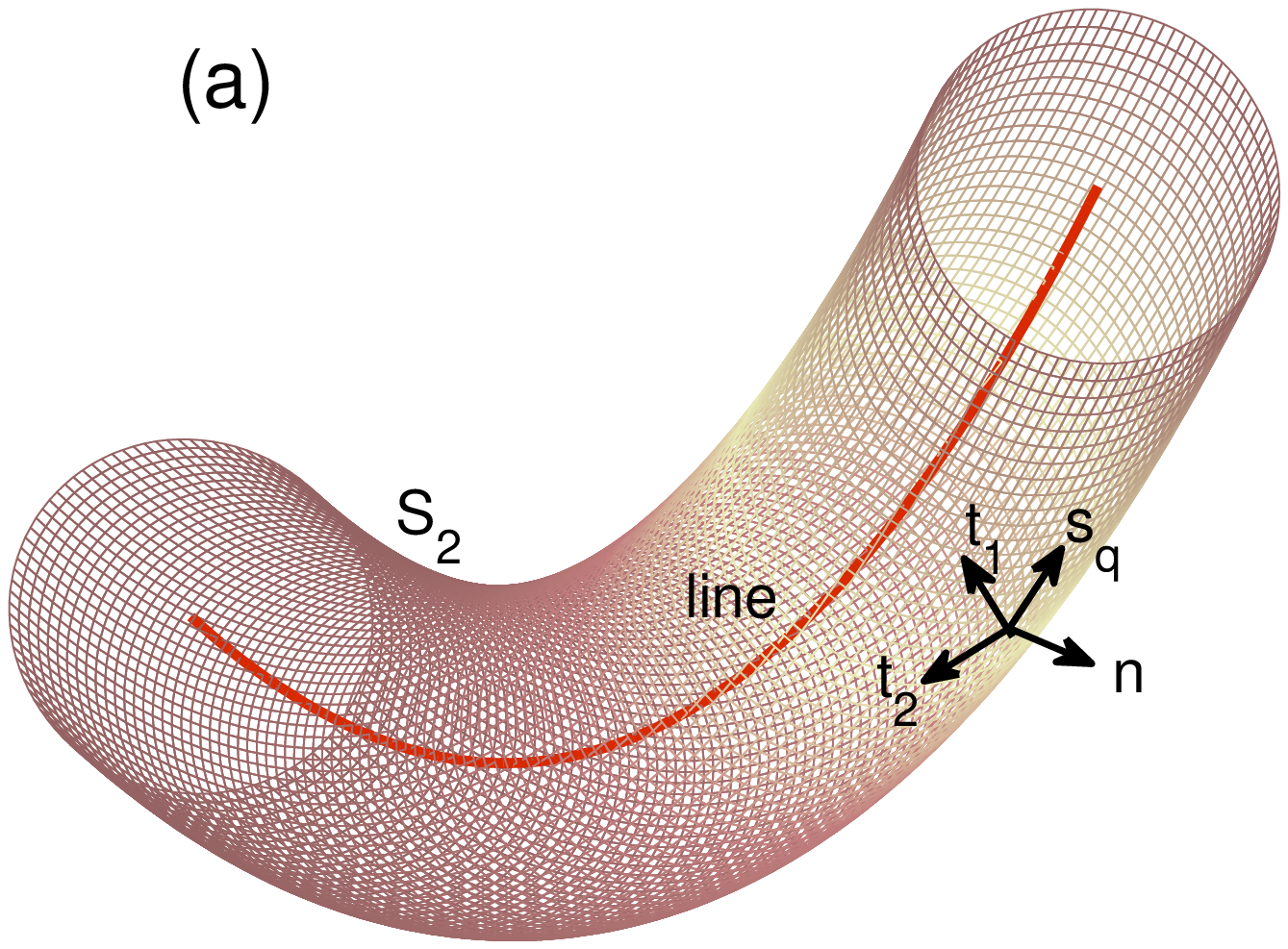}
\includegraphics[width=6cm]{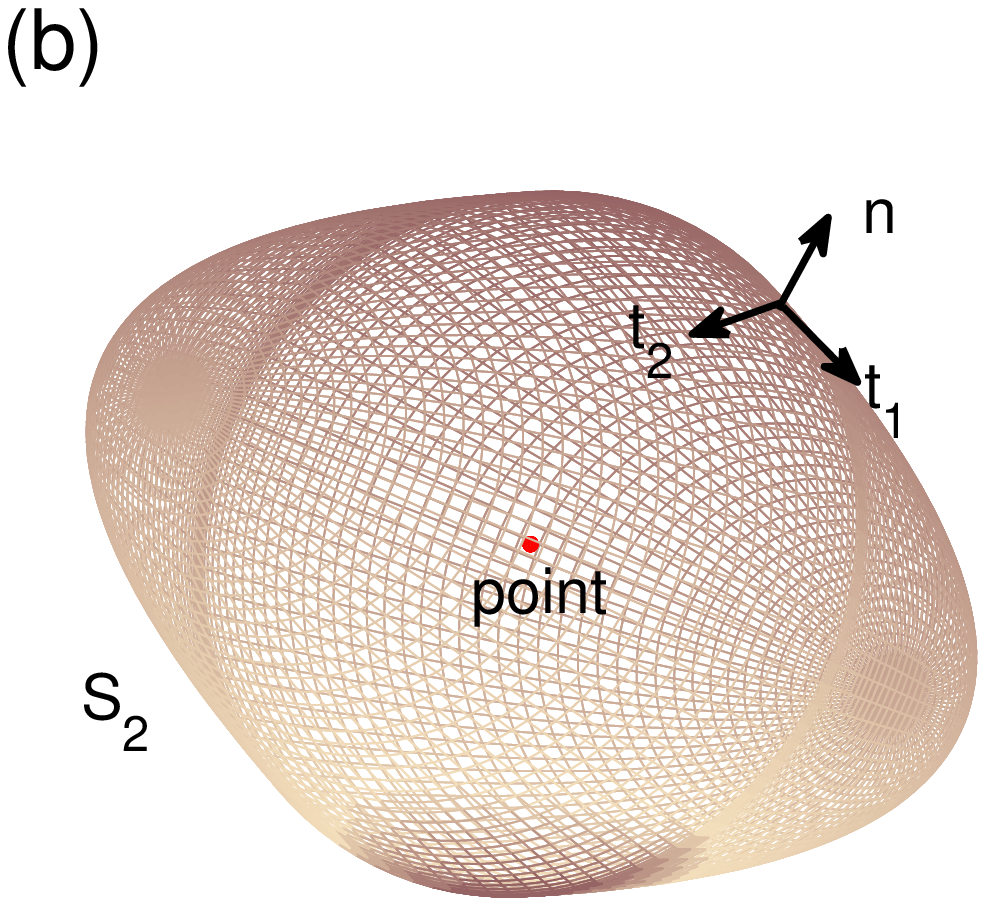}
\caption{Illustration of cloak's inner boundary for (a) a line
transformed cloak; (b) a point transformed cloak.}
\end{figure}
For the ease of our discussion, we again decompose the incident
fields at the inner boundary as $\widehat {\bf E}^{i}=[\widehat
E_n^{i} ,\widehat E_{t_1}^{i} ,\widehat E_{t_2}^{i} ]$, $\widehat
{\bf H}^{i}=[\widehat H_n^{i} ,\widehat H_{t_1}^{i} ,\widehat
H_{t_2}^{i}]$, where the subscripts $"n"$ denotes $S_2$'s normal
direction, which points outward from the cloaked region; $"t_1"$ and
$"t_2"$ denote the tangential directions of $S_2$, with $t_1$
vertical with the plane determined by two vectors in $n$ and
$\nabla_q s$ directions, and $t_2$ vertical with $t_1$. Since $s$
varies on the surface $S_2$, the direction of $\nabla_q s$ denoted
by $s_q$ should not be parallel to $S_2$'s normal direction. Thus,
the plane determined by the vectors in $n$ and $\nabla_q s$
directions always exists. The $n$, $t_1$, and $t_2$ directions are
unique, as illustrated in Fig. 2(a). $\widehat {\bf E}^{i}$ and
$\widehat {\bf H}^{i}$ at $S_2$ can also be expressed in Eq. (12),
however, with the different definitions of $n$, $t_1$ and $t_2$.
Since $f_i(q_1,q_2,z)-b_i(s)=0$ ($i=1,2,3$) characterize the inner
boundary $S_2$, $\nabla_q f_i- \nabla_q b_i $ characterize the
normal direction of $S_2$. Then $g$ at $S_2$ can be written as
\begin{equation}
g=[F_1\widehat n +\nabla_q b_1, \, F_2\widehat n +\nabla_q b_2,\,
F_3\widehat n +\nabla_q b_3],
\end{equation}
with
\begin{equation}
\fl|F_i|= \sqrt {(\partial f_i /\partial q_1-\partial b_i /\partial
q_1)^2 + (\partial f_i /\partial q_2-\partial b_i /\partial q_2
)^2+(\partial f_i /\partial q_3-\partial b_i /\partial q_3)^2 },
\end{equation}
where $F_i=|F_i|$ when the direction of $\nabla_q f_i- \nabla_q b_i$
is as the same as $n$ direction, and $F_i=-|F_i|$ when the direction
of $\nabla_q f_i- \nabla_q b_i$ is opposite to $n$ direction.

Notice that $t_1$ is orthogonal with both $n$ and $s_q$, where
$\widehat s$ denotes the unit the vector in the direction of
$\nabla_q s$. Substituting Eq. (18) into Eq. (12), it is easily
derive that
\begin{equation}
\widehat E_{t1} ^i=\widehat H_{t1} ^i  = 0.
\end{equation}
However, the other components of fields are not zero. In particular,
\begin{equation}
\widehat E_{t2} ^i  = (\widehat s \cdot \widehat t_2 )[B_1,\,B_2
,\,\,B_3]{\bf E}^i,
\end{equation}
\begin{equation}
\widehat H_{t2} ^i  = (\widehat s \cdot \widehat t_2 )[B_1,\,B_2
,\,\,B_3]{\bf H}^i,
\end{equation}
\begin{equation}
\widehat E_n ^i  = [F_1  + B_1(\widehat s \cdot \widehat n),\,F_2  +
B_2 (\widehat s \cdot \widehat n),\,\,F_3  + B_3 (\widehat s \cdot
\widehat n)]{\bf E}^i,\;
\end{equation}
\begin{equation}
\widehat H_n ^i  = [F_1  + B_1(\widehat s \cdot \widehat n),\,F_2  +
B_2 (\widehat s \cdot \widehat n),\,\,F_3  + B_3 (\widehat s \cdot
\widehat n)]{\bf H}^i,\;
\end{equation}

with
\begin{equation}
B_i  = \sqrt {\partial b_i /\partial q_1 ^2  + \partial b_i
/\partial q_2 ^2  + \partial b_i /\partial q_3 ^2 }.
\end{equation}

To further investigate how the waves interact with the inner
boundary, the values of the permittivity and permeability at $S_2$
are needed. Observing $g$ expressed in Eq. (18), it is easily
obtained that $g^T\widehat t_1=0$, indicating that $Q\widehat
t_1=gg^T\widehat t_1=0$. Thus one of $Q$'s eigen vectors is
$\widehat t_1$ with the eigen value $\lambda_{t_1}=0$, implying
$\det(g)=0$. Because $Q$ is a symmetry matrix, the other two eigen
vectors denoted by $\widehat a$ and $\widehat b$ should be
orthogonal to each other and in $n-t_2$ plane. The corresponding
eigen values are denoted by $\lambda_a$ and $\lambda_b$,
respectively, with
\begin{equation}
\lambda _a \lambda _b  = |\widehat n \times \widehat s|^2 |\widehat
F \times \widehat B|^2,
\end{equation}
where $\widehat F=F_1\widehat x+F_2\widehat y+F_3\widehat z$,
$\widehat B=B_1\widehat x+B_2\widehat y+B_3\widehat z$.

Since $\lambda_{t_1}\lambda_a\lambda_b=\det(Q)=\det(g)^2$,
$\lambda_{t_1}=\det(g)^2/(\lambda_a\lambda_b)$. Observing Eq. (10),
it is obtained that $Q\widehat{\overline{\overline \epsilon
}}=Q\widehat{\overline{\overline \mu }} = {\det (g)} $. Therefore,
it is known that $\widehat t_1$, $\widehat a$ and $\widehat b$ are
the principle axes of the cloaked medium at $S_2$ with $\widehat
{\overline{\overline \varepsilon }}$ and $\widehat
{\overline{\overline \mu }}$ expressed as
\begin{equation}
\widehat {\overline{\overline \varepsilon }}=\widehat
{\overline{\overline \mu }}=diag[\lambda_a\lambda_b/{\det(g)},\;
{\det(g)}/\lambda_a,\;{\det(g)}/\lambda_b],
\end{equation}
where the diagonal elements correspond to the principle axes
$\widehat t_1$,$\widehat a$, and $\widehat b$, respectively. Since
$\det(g)=0$, we have $\epsilon_a=\mu_a=\epsilon_b=\mu_b=0$,
indicating the cloaked medium at $S_2$ is isotropic in $n-t_2$
plane. Therefore, $\widehat n$ and $\widehat t_2$ can be considered
as the principle axes with
$\epsilon_n=\mu_n=\epsilon_{t_2}=\mu_{t_2}=0$. For $\epsilon_{t_1}$
and $\mu_{t_1}$, it is seen that they have infinitely large values.
Thus, the inner boundary operates similarly as a combination of the
PEC (perfect electric conductor) and PMC (perfect magnetic
conductor), which can support both electric and magnetic surface
displacement currents in $t_1$ direction\cite{Zhang, Greenleaf2,G2}.
In order to have zero reflection at $S_2$, the boundary conditions
at this PEC and PMC combined layer require that the incident
electric (magnetic) fields in $t_1$ direction and normal electric
(magnetic) displacement fields are all zero. From Eq. (20), we have
$\widehat E_{t_1}^{i}=\widehat H_{t_1}^{i}=0$. Since
$\epsilon_n=\mu_n=0$, it is obtained that $\widehat
D_{n}^{i}=\widehat B_{n}^{i}=0$. Therefore, it is achieved that no
reflection is excited at $S_2$, and $\widehat {\bf E}^{i}$ and
$\widehat {\bf H}^{i}$ expressed in Eq. (11) are just the total
fields in the cloak medium. The PEC and PMC combined layer
guarantees that no field can penetrate into the cloaked region. It
is worth noting that the induced displacement surface currents in
$t_1$ direction make $\widehat E_{t_2}^{i}$ and $\widehat
H_{t_2}^{i}$ at $S_2$ down to zero. However, $\widehat E_{t_2}^{i}$
and $\widehat H_{t_2}^{i}$ are not zero at the location approaching
$S_2$ in the cloak medium. Thus $\widehat E_{t_2}^{i}$ and $\widehat
H_{t_2}^{i}$ are discontinuous across the inner boundary $S_2$
\cite{Zhang, Greenleaf2, G2}.

\subsection{Case (2): point transformed cloaks}

In this case, a point with the coordinate $[c_1,c_2,c_3]$ maps to
the inner boundary. At $S_2$, we have $f_1(q_1,q_2,q_3)=c_1$,
$f_2(q_1,q_2,q_3)=c_2$, and $f_3(q_1,q_2,q_3)=c_3$. The incident
electric and magnetic fields at the inner boundary can be decomposed
into $ \widehat {\bf E}^i=[\widehat E_n^i ,\widehat
E_{t_1}^i,\widehat E_{t_2}] $ and $ \widehat {\bf H}^i= [\widehat
H_n^i ,\widehat H_{t_1}^i,\widehat H_{t_2}^i ]$, where the
definition of the subscript $"n"$ denotes $S_2$'s normal direction,
which direct outward from the cloaked region; $"t_1"$ and $"t_2$
represent $S_2$'s two tangential directions, which are vertical with
each other, as shown in Fig. 2(b). Consider $g$ at the inner
boundary, which can be expressed as
\begin{equation}
g=diag[F_1\widehat n ,\;\ F_2\widehat n ,\;\ F_3\widehat n],
\end{equation}
with
\begin{equation}
|F_i|= \sqrt {(\partial f_i /\partial q_1 )^2 + (\partial f_i
/\partial q_2 )^2 +(\partial f_i /\partial q_3 )^2},
\end{equation}
where $i=1,2,3$. Then substituting Eq. (27) into Eq. (11),  we
derive that at $S_2$
\begin{equation}
 \widehat E_{t_1}^i= \widehat H_{t_1}^i=0,
\end{equation}
\begin{equation}
 \widehat E_{t_2}^i= \widehat H_{t_2}^i=0,
\end{equation}
\begin{equation}
\widehat E_n ^i  = [F_1,\,\,F_2,\,\,F_3]{\bf E}^i,\; \widehat H_n ^i
= [F_1,\,\,F_2,\,\,F_3]{\bf H}^i.
\end{equation}
Unlike the case (1), in this case tangential fields are all zero,
implying that no field discontinuity exists at $S_2$.

Analyzing $Q$ similarly as in the case (1), we obtain that $\widehat
n$ is eigen vector of $Q$ with the eigen value
$\lambda_n=F_1^2+F_2^2+F_3^2$. While the other two eigen vectors are
$\widehat t_1$ and $\widehat t_2$ with the corresponding eigen
values $\lambda_{t1}=\lambda_{t2}=0$, indicating $\det(g)=0$.
Considering $\lambda_n\lambda_{t1}\lambda_{t2}=\det(g)^2$, we have
$\lambda_{t1}=\lambda_{t2}=\det(g)/\sqrt{(F_1^2+F_2^2+F_3^2)}$.
Therefore, $\widehat n$, $\widehat t_1$, and $\widehat t_2$ are
principle axes of the cloak medium at $S_1$ with $
\widehat{\overline{\overline \varepsilon } }$ and $
\widehat{\overline{\overline \mu} } $ given as
\begin{equation}
\fl\widehat{\overline{\overline \varepsilon }
}=\widehat{\overline{\overline \mu }
}=diag[\det(g)^2/(F_1^2+F^2+F^3),\;\sqrt{F_1^2+F^2+F^3},\;\sqrt{F_1^2+F^2+F^3}],
\end{equation}
where the diagonal elements are in the principle axes $\widehat n$,
$\widehat t_1$, and $\widehat t_2$, respectively. Since $\det(g)=0$,
$\epsilon_n=\mu_n=0$. Considering that $\epsilon_n=\mu_n=0$ and
tangential components of incident fields at $S_2$ are zero, it can
be conclude that no reflection is excited at $S_2$, and no field
penetrates into the cloaked region. The fields expressed in Eq. (11)
are the total fields in the cloak medium.

In the above sections, it has been proved that no reflection is
excited at both the exterior boundary and the inner boundary of the
cloak, and no field can penetrate into the cloaked region.
Therefore, the invisibility of invisibility cloaks with arbitrary
shape constructed by general coordinate transformations is
confirmed.

%for a 2D transformed cloak, the components of the permittivity and
%permeability in the inner boundary's tangential direction of the
%transformed plane are always infinity large, while the other
%components are zero. While for a 3D cloak, the components of the
%permittivity and permeability are all zero except the components in
%the inner boundary's normal direction.
%\section{Example}

\section{Transformation under arbitrary coordinate system}
The cloak parameters and the fields inside the cloak are expressed
in Eqs. (10) and (11), respectively. These results are expressed
under the Cartesian coordinate system, i.e., $(q_1,q_2,q_3)$
representing Cartesian coordinates in the transformed space. However
sometimes, it is much easier to discuss cloaks under other
coordinate systems, such as the cylindrical cloak under the
cylindrical coordinate system $(r,\phi,z)$. Thus, it is necessary to
obtain the corresponding expressions for cloak parameters and the
fields in a cloak under an arbitrary coordinate system, which has
also been discussed in Ref. [13].

Consider coordinate transformation, where $(q_1,q_2,q_3)$ denotes
the coordinates of an arbitrary coordinate system $(u,v,w)$ in the
transformed space. The spatial metric tensor of such coordinate
system is $Q_u=g_ug_u^T$, where $g_u$ can be obtained easily by
considering the relationship between Cartesian coordinate system and
this arbitrary coordinate system. For the cylindrical coordinate
system and spherical coordinate system, the metric tensors are
$diag[1,r^2,1]$ and $diag[1,r^2,r^2sin^2\theta]$, respectively. The
spatial metric tensor of $(q_1,q_2,q_3)$ is expressed as
$Q_q=g_qg_q^T$, where $g_q$ is shown in Eq. (4). Assuming that
$(x_1,y_1,z_1)$ are the corresponding Cartesian coordinates of
$(q_1,q_2,q_3)$ in the transformed space, the spatial metric tensor
of $(x_1,y_1,z_1)$ is then obtained as $Q_c=g_cg_c^t$, where
$g_c=g_{u1}^{-1}g_q$, and $g_{u1}$ represents $g_u$ expressed under
the coordinates $(q_1,q_2,q_3)$. Thus, from Eq. (10), the
permittivity and permeability of the cloak in Cartesian coordinates
are obtained as $ \widehat{\overline{\overline \varepsilon } } =
\widehat{\overline{\overline \mu } } = {\frac{\det (g_q)}{\det
(g_{u1})}}g_{u1}^TQ_q^{-1}g_{u1}$. Then expressing $
\widehat{\overline{\overline \varepsilon } }$ and
$\widehat{\overline{\overline \mu } }$ in $(u,v,w)$ coordinate
system, we easily have
\begin{equation}
\widehat{\overline{\overline \varepsilon } } =
\widehat{\overline{\overline \mu } }= \frac{{\det (g_q)}}{{\det
(g_{u1} )}}P_1Q_q^{-1}Q_{u1} P_1^{ - 1},
\end{equation}
where $P_1=diag[{p_1}^1,{p_2}^1,{p_3}^1]$ with ${p_i}^1=\sqrt
{g_{u1_{{\kern 1pt} i1} } ^2  + g_{u1_{{\kern 1pt} i2} } ^2  +
g_{u1_{{\kern 1pt} i3} } ^2 }$, and $Q_{u1}=g_{u1}g_{u1}^T$. As an
example, $P_1=diag[1,r,1]$ for the cylindrical coordinate system.

Consider the fields in the cloak. It is easy to know that the fields
expressed in Cartesian coordinate system are $\widehat {\bf
E}=g_c{\bf E}^i$ and $ \widehat{\bf H}=g_c{\bf H}^i$. Thus, the
fields expressed in the $(u,v,w)$ coordinate system are expressed as
following
\begin{equation}
\widehat {\bf E}=P_1Q_{u1}^{-1}g_q(g_{u0})^TP_0^{-1}{{\bf E}^i}^{'}$
, $ \widehat{\bf H}=P_1Q_{u1}^{-1}g_q(g_{u0})^TP_0^{-1}{{\bf
H}^i}^{'},\end{equation} where $g_{u0}$ represents $g_{u}$ expressed
under the coordinates $(q_1^{'},q_2^{'},q_3^{'})$ and
$Q_{u0}=g_{u0}g_{u0}^T$, $P_0=diag[{p_1}^0,{p_2}^0,{p_3}^0]$ with
${p_i}^0=\sqrt {g_{u0_{{\kern 1pt} i1} } ^2  + g_{u0_{{\kern 1pt}
i2} } ^2  + g_{u0_{{\kern 1pt} i3} } ^2 }$; ${{\bf E}^i}^{'}$ and
${{\bf H}^i}^{'}$ represent incident electrical and magnetic field
vectors expressed under the $(u,v,w)$ coordinate system.

If $(q_1^{'},q_2^{'},q_3^{'})$ denotes the corresponding coordinates
of the coordinate system $(u,v,w)$ in the original space, then $g_q$
can be written as $g_q=g_sg_{u0}$, where
\begin{eqnarray} g_s= \left[ {\begin{array}{*{20}c}
   {\frac{{\partial q_1^{'} }}{{\partial q_1 }}} & {\frac{{\partial q_2^{'} }}{{\partial q_1 }}} & {\frac{{\partial q_3^{'} }}{{\partial q_1 }}}  \\
   {\frac{{\partial q_1^{'} }}{{\partial q_2 }}} & {\frac{{\partial q_2^{'} }}{{\partial q_2 }}} & {\frac{{\partial q_3^{'} }}{{\partial q_2 }}}  \\
   {\frac{{\partial q_1^{'} }}{{\partial q_3 }}} & {\frac{{\partial q_2^{'} }}{{\partial q_3 }}} & {\frac{{\partial q_3^{'} }}{{\partial q_3 }}}  \\
\end{array}} \right].
\end{eqnarray}
Then Eqs. (34) and (35) can be expressed as
\begin{equation}
\widehat{\overline{\overline \varepsilon } } =
\widehat{\overline{\overline \mu } }= \frac{{\det
(g_s)\det(g_{u0})}}{{\det (g_{u1}
)}}P_1(g_s^T)^{-1}Q_{u0}^{-1}g_s^{-1}Q_{u1} P_1^{ - 1},
\end{equation}
\begin{equation}
\widehat {\bf E}=P_1Q_{u1}^{-1}g_s(Q_{u0})^TP_0^{-1}{{\bf E}^i}^{'}$
, $ \widehat{\bf H}=P_1Q_{u1}^{-1}g_s(Q_{u0})^TP_0^{-1}{{\bf
H}^i}^{'}.
\end{equation}

\section{Examples: cylindrical and spherical cloaks}
In this section, based on the results obtained above, the well known
radially transformed cylindrical and spherical cloaks will be
discussed as examples.
\subsection{Cylindrical cloaks}
A two-dimensional cylindrical cloak is constructed by compressing EM
fields in a cylindrical region $r^{'}<b$ into a concentric
cylindrical shell $a<r<b$. Its inner boundary is blown up by a
straight line. Thus, a cylindrical cloak is actually a line
transformed cloak. Here consider a generalized coordinate
transformation that $r^{'}=f(r)$ with $f(a)=0$ and $f(b)=b$, while
$\theta$ and $z$ are kept unchanged. Thus, $Q_{u0}$ and $Q_{u1}$
defined in the above section are $diag[1,f(r)^2,1]$ and
$diag[1,r^2,1]$, respectively, which indicates that
$\det(g_{u0})=f(r)$ and $\det(g_{u1})=r$. $P_0$ and $P_1$ are
$diag[1,f(r),1]$ and $diag[1,r,1]$, respectively. $g_s$ is equal to
$diag[f^{'}(r),1,1]$. Substituting these expressions into Eq. (37),
the permittivity and permeability of the cloak expressed in
cylindrical coordinate system are obtained easily
\begin{equation}
\epsilon _r  = \mu _r  = \frac{{f(r)}}{{rf^{'} (r)}},\epsilon
_\theta = \mu _\theta   = \frac{{rf^{'} (r)}}{{f(r)}},\epsilon _z =
\mu _z = \frac{{f(r)f^{'} (r)}}{r},
\end{equation}
It is seen that at the exterior boundary $r=b$, the cloak medium has
the PML form with $\epsilon_r=\mu_r=1/f^{'}(b)$ and
$\epsilon_\theta=\mu_\theta=\epsilon_z=\mu_z=f^{'}(b)$, which
results from the symmetry of $g_c$, which can be calculated easily.

Consider the fields $ {\bf E}^{^i }$ and $ {\bf H}^{^i }$ incident
upon the cloak. It is derived that
$P_1Q_{u1}^{-1}g_s(Q_{u0})^TP_0^{-1}=diag[f^{'}(r),f(r)/r,1]$. Then,
the fields in the cloaked medium can be obtained directly from Eq.
(38)
\begin{equation}
\fl\quad\quad\quad\widehat E_r (r,\theta ,z) = f^{'} (r)E_r ^i
(f(r),\theta ,z),\; \widehat H_r (r,\theta ,z) = f^{'} (r)H_r ^i
(f(r),\theta ,z),
\end{equation}
\begin{equation}
\fl\quad\quad\quad\widehat E_\theta  (r,\theta ,z) =
\frac{{f(r)}}{r}E_\theta ^i (f(r),\theta ,z),\;\widehat H_\theta
(r,\theta ,z) = \frac{{f(r)}}{r}H_\theta  ^i (f(r),\theta ,z),
\end{equation}
\begin{equation}
\fl\quad\quad\quad\widehat E_z (r,\theta ,z) = E_z ^i (f(r),\theta
,z),\; \widehat H_z (r,\theta ,z) = H_z ^i (f(r),\theta ,z),
\end{equation}
where $[E_r ^i,E_\theta, E_z]$ and $[H_r ^i,H_\theta, H_z]$ are
components of the incident fields expressed in cylindrical
coordinate system. When $f(r)=b(r-a)/(b-a)$, substituting the
expression of $f(r)$ into Eqs. (40)-(42), we obtain the fields in
the cloak medium, which is just the result in Ref. [6].

At the inner boundary, one can easily see that $s_q$ and $t_2$ are
both in the $z$ direction, $t_1$ is in the $\theta$ direction, and
$n$ is in the $r$ direction. Therefore, no matter what $f(r)$ is,
$\epsilon_\theta$ and $\mu_\theta$ are infinitely large, and other
components are zero, as seen in Eq. (39). $E_\theta$ and $H_\theta$,
$D_r$ and $B_r$ are all zero at $S_2$, which guarantees that no
reflection is excited at $S_2$ as analyzed above. The surface
displacement currents are induced to make $E_z$ and $H_z$ down to
zero at $S_2$. While $E_z$ and $H_z$ are not zero at the locations
approaching $S_2$ in the cloak medium. Thus, $E_z$ and $H_z$ are
discontinuous across the inner boundary \cite{Zhang, Greenleaf2,
G2}.

\subsection{Spherical Cloaks}
A three-dimensional spherical cloak can be constructed by
compressing EM fields in a spherical region $r^{'}<b$ into a
spherical shell $a<r<b$. Its inner boundary is blown up by a point.
Thus, a spherical cloak is actually a point transformed cloak. Here
a generalized radial coordinate transformation that $r^{'}=f(r)$
with $f(a)=0$ and $f(b)=b$, is considered. Similarly as in the above
example of the cylindrical cloak, the permittivity and permeability
of the spherical cloak expressed in spherical coordinate system are
derived
\begin{eqnarray}
\epsilon _r  = \mu _r  = \frac{{f(r)^2}}{{r^2f^{'} (r)}},\epsilon
_\theta = \mu _\theta   = \epsilon _\phi = \mu _\phi = f^{'}(r).
\end{eqnarray}
At the exterior boundary $r=b$, the cloak medium has the PML form
with $\epsilon_r=\mu_r=1/f^{'}(b)$ and
$\epsilon_\theta=\mu_\theta=\epsilon_\phi=\mu_\phi =f^{'}(b)$, due
to the symmetry of $g_c$.

Consider the incident fields $ {\bf E}^{^i } = [E_r^i ,\;E_{\theta
,}^i \;E_{\phi}^i ]$ and $ {\bf H}^{^i }  = [H_r^i ,\;H_{\theta ,}^i
\;H_{\phi}^i ]$ incident upon the cloak, from Eq. (39), the fields
in the cloaked medium are obtained directly
\begin{equation}
\widehat E_r (r,\theta,\phi) = f^{'} (r)E_r ^i (f(r),\theta
,\phi),\; \widehat H_r (r,\theta ,\phi) = f^{'} (r)H_r ^i
(f(r),\theta ,\phi),
\end{equation}
\begin{equation}
\widehat E_\theta  (r,\theta ,\phi) = \frac{{f(r)}}{r}E_\theta  ^i
(f(r),\theta ,\phi),\;\widehat H_\theta  (r,\theta ,\phi) =
\frac{{f(r)}}{r}H_\theta  ^i (f(r),\theta ,\phi),
\end{equation}
\begin{equation}
\widehat E_\phi  (r,\theta ,\phi) = \frac{{f(r)}}{r}E_\phi  ^i
(f(r),\theta ,\phi),\;\widehat H_\phi  (r,\theta ,\phi) =
\frac{{f(r)}}{r}H_\phi  ^i (f(r),\theta ,\phi).
\end{equation}
When $f(r)=b(r-a)/(b-a)$, the fields obtained from the above
equations agrees with the results in Ref. [5]. However, the process
of the calculation here is simpler.

At the inner boundary, observing Eqs. (45) and (46), the tangential
components of fields are zero \cite{Chen, G2, Weber}. Combining with
$\epsilon_n=\mu_n=0$, it is known that no field can penetrate into
the cloaked region.

\section{Conclusions}
In this paper, we have studied the properties of invisibility cloaks
constructed by general coordinate transformations. The invisibility
of cloaks is confirmed, by proving that no reflection is excited at
both the exterior and interior boundaries of the cloak, and no field
can penetrate into the cloaked region. The fields in the cloak
medium are related to the fields in original EM space through
$\widehat {\bf E}=g{\bf E}^i$ and $ \widehat{\bf H}=g{\bf H}^i$.
Therefore, to calculate fields in the cloak medium, there is no need
to process tedious calculations from the complex material
parameters. At the exterior boundary, when $g$ is a symmetry matrix,
the permittivity and permeability of the cloak medium have the PML
form, which is just the case for our well known radially transformed
cylindrical and spherical cloaks. At the interior boundary, the
properties of the cloak for line and point transformed invisibility
cloaks are quite different. For a line transformed cloak, the
components of the permittivity and permeability in $t_1$ direction
(defined in section V) are infinitely large. While the other
components are all zero. The fields in $t_2$ direction are
discontinuous across the inner boundary. The surface displacement
currents in $t_1$ direction are induced to make this discontinuity
self-consistent. For any point transformed cloak, at the inner
boundary, the components of the permittivity and permeability don't
have infinitely large component, and the permittivity and
permeability in the normal direction are zero. The tangential fields
at the inner boundary are zero, implying no discontinuity exist.
Therefore, comparing to line transformed cloaks, point transformed
cloaks are more practical due to the absence of the singularity of
the cloak medium.

\section*{Acknowledgements} This work is supported by the Swedish Foundation for Strategic Research (SSF)
through the Future Research Leaders program, the SSF Strategic
Research Center in Photon- ics, and the Swedish Research Council
(VR).


\section*{References}
\begin{thebibliography}{10}
\bibitem{Pendry} J. B. Pendry, D. Schurig, and D. R. Smith 2006 {\it Science}
{\bf312} 1780
\bibitem{Leonhardt} U. Leonhardt 2006 {\it Science} {\bf312} 1777
\bibitem{Cummer} S. A. Cummer, B. I. Popa, D. Schurig, D. R. Smith,
and J. B. Pendry 2006 {\it Phys. Rev. E.} {\bf74} 036621
\bibitem{Zolla}F. Zolla, S. Guenneau, A. Nicolet, and J. B. Pendry 2007 {\it Opt. Lett.} {\bf32}, 1069
\bibitem{Chen}H. S. Chen, B. I. Wu, B. L. Zhang, and
J. A. Kong 2007 {\it Phys. Rev. Lett.} {\bf99} 063903
\bibitem{Ruan}  Z. C. Ruan, M. Yan, C. W. Neff, and M. Qiu 2007 {\it Phys. Rev. Lett.} {\bf99} 113903
\bibitem{Schurig}D. Schurig, J. J. Mock, B. J. Justice, S. A. Cummer, J. B.
Pendry, A. F. Starr, and D. R. Smith 2006 {\it Science} {\bf314} 977
\bibitem{Zhang} B. L. Zhang,
H. S. Chen, B. I. Wu, Y. Luo, L. X. Ran, and J. A. Kong 2007 {\it
Phys. Rev. B} {\bf76} 121101
\bibitem{Rahm} M. Rahm, D. Schurig, D. A.
Roberts, S. A. Cummer, D. R. Smith and J. B. Pendry 2007 http:
//www.arXiv:0706.2452v1[physics.optics]
\bibitem{Ychen} H. Y. Chen and C. T. Chan 2007 {\it Appl. Phys. Lett.} {\bf90} 241105
\bibitem{Greenleaf} A. Greenleaf, Y. Kurylev, M. Lassas, and G.
Uhlmann 2007 {\it Phys. Rev. Lett.} {\bf 99} 183901
\bibitem{Greenleaf2} A. Greenleaf, Y. Kurylev, M. Lassas and G.
Uhlmann 2007 {\it Opt. Express} {\bf 15} 12717
\bibitem{U2} U. Leonhardt, T. G. Philbin 2006 {\it New J. Phys.} {\bf8} 247
\bibitem{U21} U. Leonhardt 2006 {\it New J. Phys.} {\bf8} 118
\bibitem{G2} A. Greenleaf, Y. Kurylev, M. Lassas and G.
Uhlmann 2007 {\it Comm. Math. Phys.} {\bf 275} 749
\bibitem{Weber} R. Webber 2007 http:
//www.arXiv:0711.0507[physics.optics]
\end{thebibliography}
\end{document}